\documentclass[final,twocolumn,3p,times]{elsarticle}

 \usepackage{caption}
 \usepackage{graphicx}
 \usepackage{subfig}
 \usepackage{tabularx}
 \usepackage{latexsym}
 \usepackage{amstext,amsfonts,amssymb,amsmath}
\usepackage{todonotes}
\usepackage[version=3]{mhchem}		% chemistry formulas

\usepackage{epstopdf}
\usepackage{multicol}
\usepackage{amsmath}
\usepackage{nonfloat}
\usepackage{color}
\usepackage{bm}

\usepackage{jabbrv}
\biboptions{sort&compress}

\newcommand{\EQ}{\begin{equation}}
\newcommand{\EN}{\end{equation}}
\newcommand{\NC}{\begin{numcases}{}}
\newcommand{\NN}{\end{numcases}}
\newcommand{\EQA}{\begin{eqnarray}}
\newcommand{\ENA}{\end{eqnarray}}

\biboptions{comma,sort&compress}
\begin{document}

\begin{frontmatter}
\title{Emerging Trends in Numerical Simulations of Combustion Systems}
\author[label1]{Venkat Raman}
\author[label1]{Malik Hassanaly}
\address[label1]{Department of Aerospace Engineering, University of Michigan, Ann Arbor, MI 48109, USA}

\begin{abstract}
Numerical simulations have played a vital role in the design of modern combustion systems. Over the last two decades, the focus of research has been on the development of the large eddy simulation (LES) approach, which leveraged the vast increase in computing power to dramatically improve predictive accuracy. Even with the anticipated increase in supercomputing capabilities, the use of LES in design is limited by its high computational cost. Moreover, to aid decision making, such LES computations have to be augmented to estimate underlying uncertainties in simulation components. At the same time, other changes are happening across industries that build or use combustion devices. While efficiency and emissions reduction are still the primary design objectives, reducing cost of operation by optimizing maintenance and repair is becoming an important segment of the enterprise. This latter quest is aided by the digitization of combustors, which allows collection and storage of operational data from a host of sensors over a fleet of devices. Moreover, several levels of computing including low-power hardware present on board the combustion systems are becoming available. Such large data sets create unique opportunities for design and maintenance if appropriate numerical tools are made available. As LES revolutionized computing-guided design by leveraging supercomputing, a new generation of numerical approaches is needed to utilize this vast amount of data and the varied nature of computing hardware. In this article, a review of emerging computational approaches for this heterogeneous data-driven environment is provided. A case is made that new but unconventional opportunities for physics-based combustion modeling exist in this realm.

\begin{keyword}
Large eddy simulations \sep Uncertainty quantification \sep Reduced-order models \sep Data driven models \sep Digital twins \sep Rare events
\end{keyword}

\end{abstract}

\end{frontmatter}

\clearpage
\setcounter{page}{1}
\section{Introduction: Trends in combustion modeling applications}
\label{sec:intro}
Numerical computation of turbulent flames has become a key tool in the quest for efficient and low-emission combustors \cite{veynante_review,haworth_review,echeckki_book}. While empirical and/or analytical models enabled much of the earlier progress in understanding turbulent flames \cite{borghi_review}, the modern approach relies on the numerical solution of reduced forms of the conservation equations for mass, momentum, and energy. Broadly termed, computational fluid dynamics (CFD) refers to both the development of the reduced equations and their numerical resolution for appropriate configurations. It has long been established that turbulent combustion spans a large range of length and time-scales, which makes capturing the entirety of the combustion process an intractable computational problem for realistic flows. As a result, models that reduce this computational complexity are necessary for the simulation of practical flows. 

Although a number of different modeling approaches have been developed over the last few decades, there have been two primary frameworks, based on the level of description of the underlying turbulent flow: the Reynolds-averaged Navier Stokes (RANS) and the more recent large eddy simulation (LES) approaches. Of these, LES has become the de-facto numerical tool due to its ability to accurately model {turbulent physical processes that are relevant to combustion applications}~\cite{pitsch_arfm}. Further, the LES framework is able to leverage the growth in computing power by progressively increasing the range of physical length and time scales that are directly resolved rather than modeled. In this sense, LES provides a natural bridge to nearly model-free direct numerical simulations (DNS) \cite{jackie_review}. As a result, model development in the turbulent combustion community has almost exclusively focused on the LES approach, with progress in many different areas \cite{benoit_review,menon_pecs,haworth_review,jenny_review,ramanfox_arfm}.

As LES models evolve and mature, it is necessary to assess their role in the context of end applications. So far, the main utility of computational models has been in the design of combustors for a variety of applications. In particular, the focus has been on flame processes and emissions (NOx, soot, unburnt hydrocarbons). A typical use is to simulate specific designs to evaluate emissions or temperature profiles at the exit of the combustor. This information is then used by an expert to make modifications to the design based on the end goals. Currently, the number of such simulations used in a design cycle is mostly limited by the computational expense of individual runs \cite[chap. 29]{witherTurbulenceBook}. In this context, the computational power has led to an increase in model complexity, so much so that routine use of such high-fidelity computational models is not practical when rapid turnaround of computations is necessary. A second issue is that combustors are only one component of the propulsion or energy-conversion devices, and interact with turbomachinery or other components located upstream or downstream in the flow path. Since CFD-based models solve partial differential equations (PDEs), and PDEs are sensitive to boundary conditions, there is an inherent uncertainty in the results of the computations. Even when models are accurate, their reliability is only valid for certain flow conditions. As a result, they may introduce errors when used outside these regimes of validity. In the design context, not only are the results important but also is their sensitivity to these uncertainties. Hence, even with the remarkable growth in computational power and the successes in LES modeling, the direct use of such tools for design remains a challenge. 

Across society, concepts such as autonomy and digitization and digitalization are becoming prevalent, changing everything from thermostats to aircrafts. Digitization is {the process of using sensors and other measurements devices to collect data in a digital format. In the industrial context, it refers to instrumenting devices in a way that measurements can be accessed digitally}. Digitalization leverages digitization to improve performance or any other metric of interest. In the last few years, the gas turbine sector has increased its focus on strict cost control as a way to provide value to its customers \cite{dtgt2}. A significant driver of cost is the maintenance, repair and overhaul (MRO) process. Optimizing the maintenance process through reduced down-time is therefore a key cost-saving measure. While a large share of this maintenance issue is related to the structural components of the gas turbine, the combustion processes drive the thermal loads. Hence, the ability to use advanced CFD tools for forecasting performance issues is beginning to be a critical industry need. Since these tools will have to be used not for a single device but a fleet of gas turbines, computational cost is the most important constraint. A connected development is the digitalization of manufacturing itself, which allows engine components to be available as digital copies that can be readily used to conduct computational studies. To manage this complex MRO process, industry has developed a host of computational methods derived from the CFD models, such as ``component zooming" \cite{pachidis_asme}, multi-level modeling \cite{dtgt2}, and digital twinning \cite{kurzke,visser}. However, these tools are currently only loosely connected to the physics-driven computational modeling research. 

These discussions about computing cannot be complete without the inclusion of data sciences. Much like the last two decades were dominated by high performance computing, the next decade (at the least) is poised to be the data era. While early progress in the use of data was in fields that did not possess an underlying physical basis (social media, linguistics), the use of data is now spreading to physical sciences as well \cite{kutz_book}. Within turbulent combustion, diagnostic tools have made as much progress as computing has in the last few decades, leading to a wealth of data even in extreme operating conditions (high temperature, high pressure and complex geometries). Even if such data can be generated and stored, its dissemination is an important challenge \cite[chap. 28]{witherTurbulenceBook}, which requires a community effort.  While such data is often used for validation of models, the advances in data sciences indicate a potential for innovation. Further, practical gas turbines are increasingly instrumented with a host of sensors that are used not only for operational control, but also for capturing and transmitting data at varying rates to centralized repositories \cite{remote_gas}. As a result, tools used for data mining and machine learning such as artificial neural networks can now be used to deal with fault diagnosis \cite{zedda_aiaa,ogaji_riti_ieee} or even forecasting \cite{sarkar_symbolic}. However, such purely data-driven approaches without a constraining physical model cannot be reliable, especially when forecasting operational characteristics at conditions not present in the data used to obtain the models.

The discussion above shows that a) there is a critical gap in translating state-of-the-art computational models into a design or analysis tool, and b) there exist new opportunities for expanding the scope of combustion modeling beyond the CFD-driven sub-grid/sub-filter modeling. The success of LES was due to its ability to utilize supercomputing in order to advance predictive capability. In the same vein, the next generation of combustion modeling will have to infuse other emerging elements, including the changing application needs and the availability of data. The review below seeks to introduce possible paths for numerical simulations and key challenges. It will be seen that there is already pioneering work from the combustion community in some of these areas. The rest of the article is organized as follows. Section~\ref{sec:current} discusses the advances in LES and the current challenges in the post-design scope of combustion modeling. Section~\ref{sec:uq} provides a review of uncertainty quantification as the starting point for data infusion. Section~\ref{sec:roms} provides information on data-driven modeling. Section~\ref{sec:future} provides sample applications that utilize the tools discussed in the preceding sections, and also identifies areas that need additional breakthroughs. Finally, an outlook section (Sec.~\ref{sec:outlook}) summarizes these topics and provides a few conclusions.

\section{Current status and modeling trends}
\label{sec:current}
Numerical modeling of turbulent combustion usually implies conducting LES or RANS simulations, with the former dominating the research focus of the community. In order to discuss emerging trends, it is useful to take stock of the enormous advances that LES has provided. The scope of this section is not to discuss every facet of modeling, for which the reader is referred to other recent reviews \cite{pitsch_arfm,haworth_pecs,benoit_review,menon_pecs}. Rather, the discussion here is to determine the state of the models that will be relevant to the expanded scope of combustion applications discussed in Sec.~\ref{sec:intro}.

In LES, the turbulent flow is separated into resolved and unresolved motions, with the resolved features directly solved using transport equations and the unresolved features are modeled using sub-filter closures. While such separation of variables is the first step for any modeling framework, the filtering operation is specifically suited for turbulent flows that exhibit a forward energy cascade. This ensures that the macroscopic flow features are governed by the large scales. The LES approach was originally formulated for atmospheric flows~\cite{smagorinsky_circulation}, then adopted for non-reacting turbulent flows \cite{mansour_70s,moin_70s}, and later extended to turbulent reacting flows~\cite{pitschSteiner,nottin,jimenez1997priori,haworth_ice,menon_ramjet_early}. While the approach initially provided theoretical promise, its success in combustion is due to critical modeling advances \cite{pitsch_arfm}.

\subsection{In-situ evaluation of model coefficients}

The first major advance in LES was the development of the dynamic modeling procedure (DMP) \cite{moin_C,germano,lilly}. One of the main shortcomings of the preceding RANS approach was the need to specify coefficient values that appear in the various constituent models. Since these coefficients need not be universal constants, simulation predictions were found to be highly sensitive to the choice of these variables \cite{raman_rans,rans_sensitivity}. The DMP obviated this need in LES by determining model coefficients on the fly, using information from the flow-field solution at a given time-step. The underlying concept here is that LES solution fields can be filtered (down-sampled) to create surrogate LES fields at larger filter widths. If model coefficients are assumed to be invariant to the filter size, then by comparing the LES and surrogate fields, these values can be extracted. For instance, DMP was used to model turbulent diffusivity \cite{pitschSteiner}, which enabled accurate prediction of radial fuel spread in canonical jet flames. Extensions to scalar variance and dissipation rate \cite{pierce_moin,balarac,knudsen,cookrileydissipation} vastly improved sub-filter closures for chemically reacting flows. {In the context turbulent premixed flames, the DMP has been used to model the sub-grid scale flame wrinkling leading to increased accuracy in capturing flame dynamics. In particular, the DMP has been used to model the flame wrinkling factor \cite{charlette_dmp}, to close the flame surface density equation \cite{knikker_dmp} and to model the level set equation \cite{knudsen_dmp}. These techniques have been recently examined and compared using a DNS database of a realistic swirl combustor \cite{veynante_dmp}.} While most of the models closed using the DMP are algebraic in nature, this procedure has been developed for transport-equation based models as well \cite{dlm_ghosal,kaul_pci,kaul_ctm}.  

The DMP can be applied to models of quantities that are predominantly governed by large scale processes. For instance, sub-filter variance or turbulent diffusivity are both controlled by gradients of the scalar and velocity fields at the filtered scale. On the other hand, predominantly small-scale quantities such as scalar dissipation rate or chemical source term, require explicit modeling since these terms are dependent on the structure of the scalar field at the smallest length scales. As a result, modeling the turbulence-chemistry interaction is important in LES just as in RANS. However, it has been found that RANS-type combustion models when used in LES lead to better results, mainly because the input to the combustion models, which depend on the large scales, are better captured \cite{pitsch_arfm}.

\subsection{Advances in combustion models}

In this context, the second major success in LES has been the development of a hierarchy of combustion models - often rooted in the RANS methodology and adapted for the filtered formulation - which have proven to be highly accurate for a diverse set of applications. Although all RANS models have been ported to LES \cite{givi_aiaa,pitsch_arfm,ces_paper,colin_atf}, tabulation-based methods are beginning to emerge as the primary LES modeling approach. The general formulation in LES can be described as a mapping:
\begin{equation}
    {\bm \phi} = {\cal G}({\bm \xi}),
\end{equation}
where ${\bm \phi}$ denotes the local thermochemical composition vector of length $N_s$, and ${\bm \xi}$ is a vector of input variables of length $N_m$ to the combustion model ${\cal G}$. Since LES provides only filtered values, the needed mapping for LES is obtained as
\begin{equation}
    \widetilde{\bm \phi} = \int {\cal G}({\bm \xi}) P({\bm \xi})d {\bm \xi},
\end{equation}
where $P$ is the sub-filter joint PDF of the input variables \cite{colucci,givi_aiaa}. The combustion modeling problem is then the specification of both ${\cal G}$ and $P$. Different methodologies tackle each of these terms to different levels of complexity. For instance, transported PDF methods solve directly for $P$ while assuming unity mapping (i.e., ${\bm \phi} = {\bm \xi}$, $N_m = N_s$). More recently, the use of tabulated approaches has gained significant interest, so much so that different tabulations have been used for nearly every combustion scenario \cite{fpva_pierce,strat_flame,partially_premixed_knudsen,spray_chrigui,soot_mueller,saghafian_scram}. Turbulent diffusion flames that are far from extinction have been successfully modeled using steady laminar flamelet models \cite{pitschSteiner,kempfslfm}. However, even when the flame structure is more complex, for instance in partially-premixed \cite{knudsen} or spray flames \cite{spray_chrigui,roekaertspapers}, the use of the flame-structure approach with appropriate enhancements have led to good reproduction of experimental data. The research focus is on determining appropriate canonical flow problems that can be used to develop the mapping function ${\cal G}$ that best represents the end application. For instance, heat-loss to walls has been modeled by using an enthalpy related mapping variable \cite{benoit_heatloss,mueller_soot,shunn_radiation,yihao_3jet}. Similarly, the issue of homogeneous reactions, for instance to describe kernel growth in an ignition case, have been modeled using a well-stirred reactor based approximation \cite{pera,yihao}. In the tabulated approach, the PDF $P$ still needs to be prescribed, and several options have been considered \cite{raman_bluffbody,mueller_PDF,doran_PDF,smld_pope, smld_ihme}. 
{In general, the modeling of turbulent premixed flames has followed a different approach, where for low Karlovitz number configurations, the reaction zone is treated as thin compared to other relevant length scales. This leads to a flame surface based approach, resulting in many different formulations \cite{pv02,peters2000turbulent}. In many practical flows, the filter width is larger than the flame thickness, which results in an under-resolved description of the surface. To overcome this problem, one strategy is to use flame thickening techniques in order to artificially fix the thickness of the flame front at the LES filter size. This can be achieved by altering the scalar diffusivity~\cite{colin_atf} or using filtered flamelet models~\cite{ftacles}. In order to accurately capture the turbulent flame propagation, the modification of the flame structure at the unresolved scales needs to be considered. Substantial efforts have been dedicated to capturing flame speed using flame wrinkling factors \cite{charlette_wrinkling}, and various closure terms for other flame description approaches \cite{chakraborty_fsdclosure,moureau_levelset}.} Apart from these techniques, other approaches also exist \cite{menon_lem,gequation}, and are used for specific applications.

\subsection{Advances in numerical algorithms}

A third and important focus in LES has been on the underlying numerical approaches. Almost all LES formulations treat the filter width as comparable to the mesh size, which creates a unique numerical problem: as the computational mesh is altered, the contribution of the sub-filter model changes due to the change in the filter width. Unlike other PDE applications, LES is not directly amenable to grid convergence. Hence, reducing numerical errors for a given mesh size has been at the forefront of LES research. In this regard, algorithms for secondary conservation (i.e., for variables that are not solved directly) have been used to ensure the robustness of LES algorithms \cite{hamiaccarino,morinishi_skew,mahesh,kravchenko,akselvoll}. Where feasible, reformulation of the underlying equations has been used to decrease numerical issues \cite{seqmom1,seqmom2,kaul_pof1,kaul_pof2}. Conceptually, the main advantage of LES over RANS is its ability to utilize supercomputing to reduce the effect of models. In other words, the filter width can be progressively decreased with an increase in computing power, leading to results that approach a model-free evolution of the governing equations. This has naturally led to porting of LES solvers to massively parallel computers \cite{moureau_solver,UQrichdome_iaccarino2017,malik_openfoampaper}. This ability to leverage large computing resources allows faster time to solution, but also allows increasingly complex sub-filter models to be used.

\begin{figure}
\center
\includegraphics[width=0.49\textwidth,trim={0cm 0cm 0cm 0cm},clip]{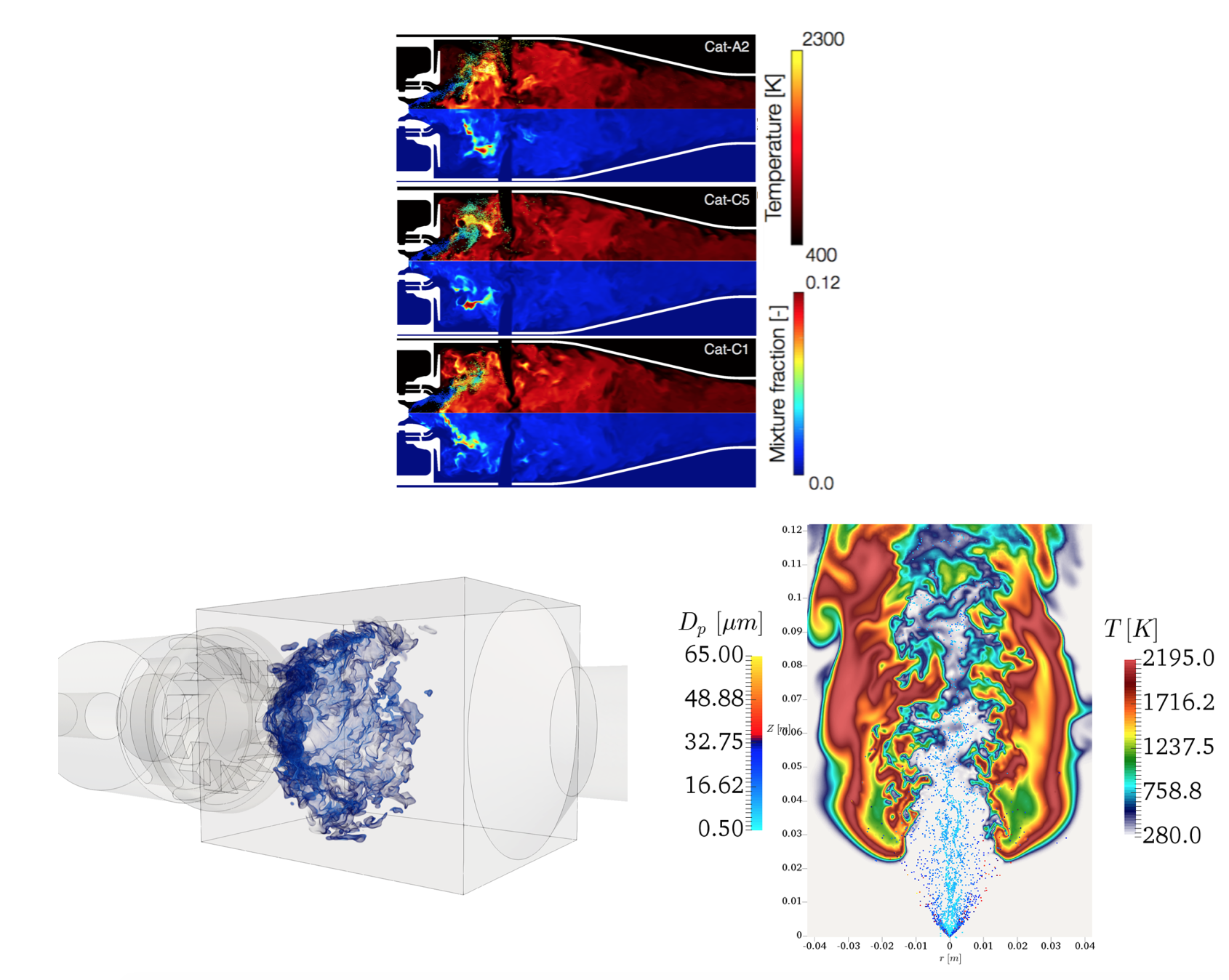}
\caption{Sample of high-fidelity simulations including complex geometry and complex flow physics. Top: simulation of the referee combustor rig, including liquid spray injection and side-jet induced recirculation. Reproduced from Ref.~\cite{ihme_figure2}. Bottom left: volume rendering of a flame isosurface in the Preccinsta burner. Based on Ref.~\cite{moureau2011large}. Bottom right: temperature contour in a turbulent spray burner. Based on Ref.~\cite{spray_moureau}.}
\label{fig:current}
\end{figure}

These three scientific pillars{- dynamic modeling procedure, advances in modeling methods and in numerical methods -} have made LES into a preeminent tool for modeling complex reacting flows. For instance, full scale simulations of combustors are now increasingly routine \cite{dlr_alex,pw2000_matthias,helicopter_poinsot,dimare_combustor,ice_pitsch,fureby_hyshot} for all types of applications. Figure~\ref{fig:current} shows a sample of such calculations including complex geometry, recirculating/swirling flows and multiphase physics. With increasing access to computing power, such computations will allow shorter time-to-solution, which will further increase their utility in the design process. However, it is important to recognize the limitations of LES: a) while LES is accurate in predicting macroscopic quantities, other pollutants such as soot face considerable modeling challenges \cite{ramanfox_arfm,mueller_pof,mueller_thesis}; b) in liquid fuel based combustion devices, the treatment of multiphase flows is still a subject of considerable research focus with limited demonstration of LES validity \cite{jenny_review,tcsworkshops,herrmann_arfm}; c) in spite of the advances in computing, LES is still expensive for design space explorations. This is especially the case for multi-injector configurations (rocket motors, annular combustors) or capturing unsteady features (precessing structures, thermoacoustics or scramjet unstart); d) even though LES is unsteady and three-dimensional, it can be shown that these computations only reproduce an average path in phase-space \cite{langford_moser,pope_self}. As a result of these issues, a single LES computation is still subject to uncertainties in operating and boundary conditions as well as in the models. Even given these limitations, in our view, the use of LES will grow in the near term, mainly due to a lack of competing modeling frameworks. 

%\begin{figure}
%\center
%\includegraphics[width=0.49\textwidth,trim={0cm 0cm 0cm 0cm},clip]{./Figures/collage.png}
%\caption{Sample of high-fidelity simulations including complex geometry and complex flow physics. Top: simulation of the referee combustor rig, including liquid spray injection and side-jet induced recirculation. Isosurface of reaction source term colored by termperature and grey isosurface of the side-jet velocity (left). Reproduced from Ref.~\cite{ihme_figure}. Instantaneous contour of temperature and mixture fraction for three different fuels (right). Reproduced from Ref.~\cite{ihme_figure2}. Bottom left: volume rendering of a flame isosurface in the Preccinsta burner. Reproduced from Ref.~\cite{..}. Bottom right: ...    }
%\end{figure}

However, there is a critical need to address the limitations identified above. Given the expanded scope for combustion modeling (Sec.~\ref{sec:intro}), LES has to be augmented or complemented by other computational tools able to answer different type of questions. The following sections provide some possible paths, focusing on different aspects of numerical simulations.

\section{Uncertainty quantification}
\label{sec:uq}
One of the utilities of computational models is to enable informed decisions on complex processes. However, simulations themselves are imperfect and subject to numerous sources of uncertainty, which degrades the usefulness of the results. Improving reliability requires estimation of these errors, which is broadly termed as uncertainty quantification (UQ). While UQ techniques have been prevalent in a number of other disciplines, their application to engineering began in earnest in the mid-1990's, predominantly driven by the US Department of Energy interest in the modeling and simulation of complex engineered systems \cite{asc}. In simple terms, UQ is the process of adding error estimates or confidence intervals to simulations. Depending on the nature of the uncertainty, the end goal can be either to provide a confidence interval on the estimate of the quantity of interest (QoI), or predict extreme scenario for the QoI. Although experimental sciences have developed a robust culture of estimating and reporting such confidence intervals for measurements, UQ is incredibly challenging for computational sciences and has not received similar attention. The UQ sciences aim to address this gap by developing both computational tools and scientific methodologies for estimating the uncertainty of models. 

\subsection{Uncertainty sources}

\begin{figure}
\center
\includegraphics[width=0.49\textwidth,trim={0cm 0cm 0cm 0cm},clip]{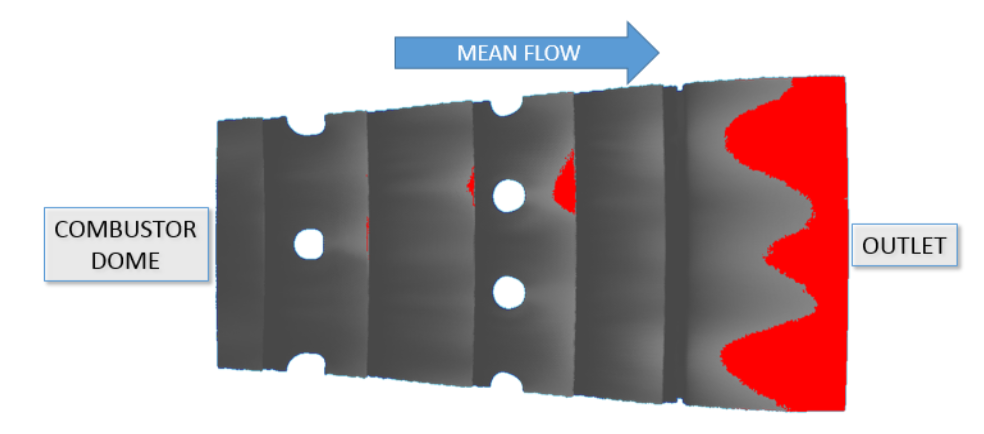}
\caption{Impact of inflow condition uncertainty on the wall heat transfer on an aircraft combustor. The figure shows the standard deviations of the time-average temperature at the wall. Grey color scale denotes standard deviations below 2 times the overall standard deviation. Red color denotes standard deviation more than 2 times the overall standard deviation. The result highlights where hot spots could occur due to uncertainty in inflow conditions. Reproduced from Ref.~\cite{UQrichdome_iaccarino2017}.}
\label{fig:richdome}
\end{figure}

There exist multiple sources of uncertainty, but the end goal is to characterize their impact on QoIs that are derived from the simulations. For turbulent reacting flows, there are two different sources of uncertainty:
\begin{itemize}
    \item \textbf{Uncertainty in simulation conditions:} This includes 1) lack of precise knowledge of initial and boundary conditions, as well as 2) the variability associated with the geometry of the combustion device. The first source of uncertainty is due to the fact that only bulk statistical properties are available from upstream and downstream gas turbine components as input to the LES calculation. For instance, Constantine et al.~\cite{ActiveSubspacesHyshotUQ} evaluate the impact of the turbulence length scale and turbulence intensity on a QoI quantifying unstart in scramjet isolators. Similarly, Masquelet et al.~\cite{UQrichdome_iaccarino2017} investigated the impact of fuel flow rate and inflow temperature on an aircraft gas turbine. Figure~\ref{fig:richdome} illustrates the part of the domain on which uncertainties have the most impact. Here, the wall temperatures of an extreme case are compared to the mean temperature plus one standard deviation obtained with the UQ procedure. They find that inlet parameters uncertainty mostly affect the combustor outlet.  In the context of ignition, uncertainty in the spark conditions can have dramatic effect on the outcome and can lead to ignition success or failure. This is of primary interest for the problem of high-altitude relight~\cite{yihao}. Mueller and Raman~\cite{effectModelErrorSootMueller} studied the impact of upstream prediction errors on soot statistics downstream in a jet flame. Geometry variability is more critical for MRO applications where device degradation may not be precisely known, leading to changes in geometric features that can impact device performance. For instance, Ref.~\cite{dataDrivenFaultDetection_ray_2008} studied the impact of friction coefficient on thermoacoustic instability. 
    \item \textbf{Uncertainty in models:} Since combustion simulations use a host of models, the inaccuracies of these models is a dominant source of uncertainty in predictions. These models errors are, in turn, of two kinds: a) parametric uncertainties resulting from non-universality of model coefficients, such as rate parameters in chemistry models \cite{sheen,StatisticalDeterministicFrenklach} or turbulence model coefficients. In LES, the impact of these parameters on predictions have been studied \cite{UQFlameletSandiaD_mueller2013,reagan_cnf,review_wang,UQModelCoefficientLES_khalil}; b) model form or structural uncertainties arising from the specific modeling assumptions made and functional forms chosen \cite{oliver_moser}. This source of error has not been widely studied in LES \cite{peer_ramanmueller2018}. In the RANS context, model errors for Reynolds stress \cite{structuraluncertaintyRANS_emory2013} or the flamelet-progress variable approach \cite{superviseLearningUQDuraisamy} have been quantified. The model form uncertainty introduces another context for modeling, which is discussed further below.
\end{itemize}

\subsection{Uncertainty representation and propagation}

When the perceived errors in model parameters are small, sensitivity tools are useful \cite{cacuci2005sensitivity,tomlinsonturanyi}. For systems with a large number of parameters, adjoint-based approaches have been used in RANS \cite{oliverDarmofal} and laminar flows \cite{Adjointbasedsensitivityanalysisofflames}. However, these tools are not readily extendable for LES, which features chaotic dynamics \cite{wang_sensitivity, jesse-adjoints} and requires specialized, computationally intensive approaches. When uncertainties are large, a probabilistic Bayesian analysis has become the preferred approach \cite{ARFMNajm}, although other non-probabilistic techniques exist \cite{StatisticalDeterministicFrenklach,hansen,obserkampfevidencetheory}.

In the Bayesian approach, the set of uncertain parameters is represented as a joint probability density function (PDF). This joint-PDF is obtained by an estimation process, where data from a hierarchy of experiments (simple canonical flows to increasingly complex flows) that can exercise specific parameters (or sub-sets of parameters) are used along with the Bayesian calibration process \cite{oliver_bayesian1,oliver_bayesian2}. Based on the data, the spread of the marginal PDF for certain parameters might decrease, representing an increase in confidence in the chosen values for those parameters. The uncertainty in the QoI is then found by an ensemble of CFD calculations, where each simulation uses a set of parameter values drawn from the joint-PDF. A brute-force Monte-Carlo approach faces the curse of dimensionality, when the number of parameters is large.

The practical use of UQ approaches {requires tackling several challenges}. From a propagation perspective, efficient techniques to propagate uncertainty of a large number of parameters through simulation tools become necessary. In this context, efficient techniques such as the polynomial chaos expansion method (PCE) \cite{ARFMNajm} have been used for fluid mechanics applications. Further, the number of parameters can be reduced by discovering strong correlations through the active-subspace approach \cite{ActiveSubspacesHyshotUQ}. Even when using these approaches, the total computational cost of performing an ensemble of LES may be prohibitive. There have been sparse implementations of such techniques such as that of Mueller et al.~\cite{UQFlameletSandiaD_mueller2013}, who used the tabulation approach for flamelet models to encapsulate the errors due to the underlying chemistry mechanism. In this way, the effect of kinetics uncertainty can be efficiently treated in LES computations.

In combustion applications, structural uncertainty in models will be the dominant source of errors \cite{peer_ramanmueller2018}. In this case, a model $M({\bm \theta})$ for a quantity $y$ which is a function of input variables ${\bm \theta}$ can be expressed as
\[
y = M({\bm \theta}) + \varepsilon,
\]
where $\varepsilon$ is the error introduced by the model. There are different approaches to estimate this error. In the purely data-driven approach, an appropriate dataset is used to calibrate $\varepsilon({\bm \theta},\beta)$, where $\beta$ are parameters associated with this error \cite{oliver_cmame}.

\begin{figure}
\center
\includegraphics[width=0.40\textwidth,trim={0.0cm 0cm 0cm 0cm},clip]{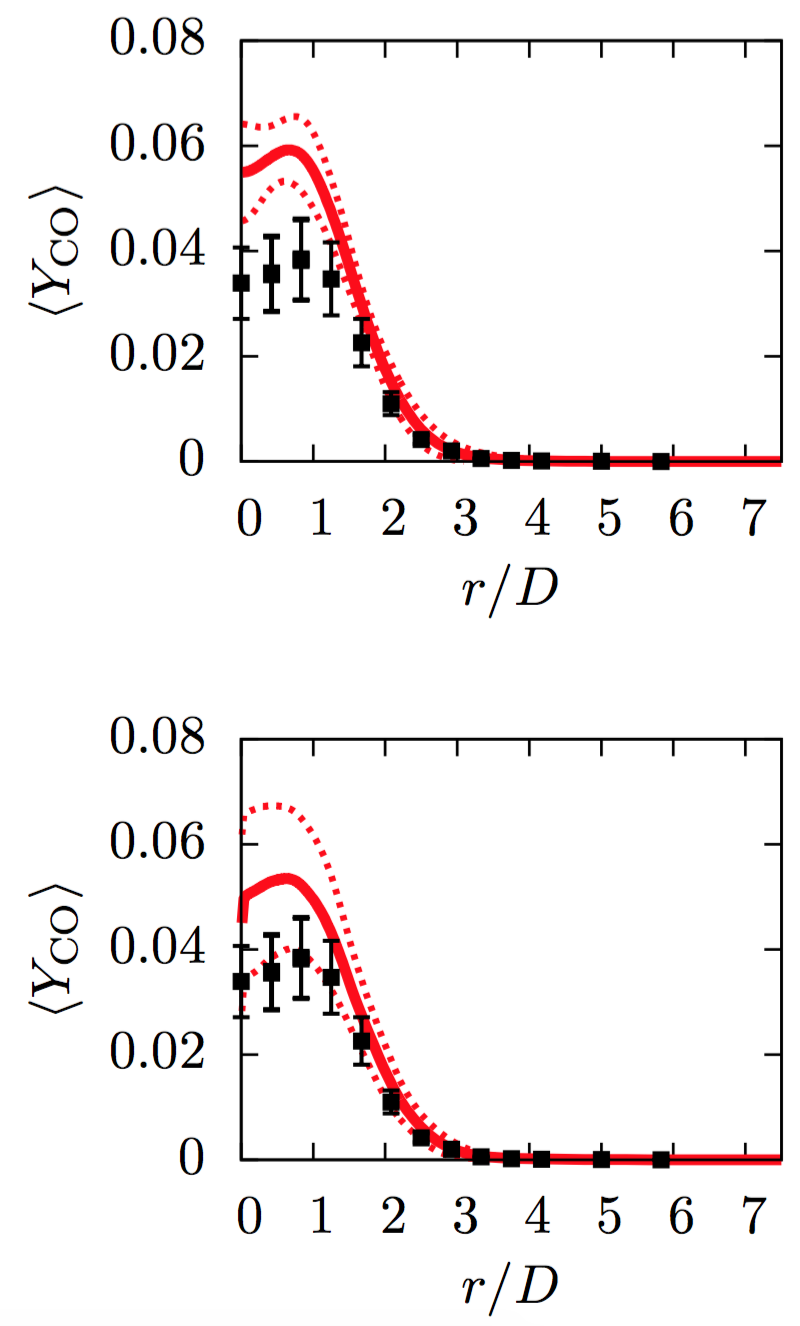}
\caption{Uncertainties of time-averaged CO mass fraction in the Sandia D flame, due to subfilter mixture fraction dissipation model (top) and chemical kinetic rates (bottom). The solid line shows the mean prediction, while the dotted lines show the one standard deviation prediction due to uncertainty. The symbols represent experimental data. Reproduced from Ref.~\cite{peer_ramanmueller2018}.}
\label{fig:peer}
\end{figure}

Alternatively, such errors can themselves be modeled: these are termed physics-based model form errors \cite{peer_ramanmueller2018}. Here, the quantitative estimates of the errors due to physical assumptions are directly modeled. Within this approach, two sub-classes are defined: a) a hierarchical approach, where higher-fidelity models are used to develop uncertainty estimates for lower fidelity models, and b) a peer-model approach where two models with disparate physical assumptions are used to quantify the variability in the simulation results. This latter approach is demonstrated in Ref.~\cite{peer_ramanmueller2018} by estimating the model uncertainty for scalar dissipation rate. The two peer models for dissipation rate that were considered include a large-scale based mixing time scale and a chemical time scale. A composite dissipation rate is formed by linearly combining the two models with an uncertain parameter $\beta$ that is assumed to vary uniformly. Figure~\ref{fig:peer} shows the spread in predictions for CO mass fractions due to this estimated uncertainty in dissipation rate models.

From these discussions, it is concluded that UQ for combustion applications is an important yet nascent field. An efficient approach for propagating the uncertainty of large numbers of parameters involved in combustion models is necessary. At the same time, the modeling of structural uncertainty opens a new opportunity for research, where the focus is on representing physical assumptions in a mathematical formulation. It is also seen that UQ provides a way for introducing data (experimental or higher-fidelity simulation) into the models. This important starting point is further expanded in the following section to move towards data-driven modeling.

\section{Data-based and reduced-order modeling}
\label{sec:roms}
As discussed in the previous two sections, while high-fidelity modeling is useful, it is not a comprehensive solution for the modeling problem. Many practical systems are far too complex to envision a brute-force use of such tools. As an example, gas turbines typically contain several circumferentially arranged combustors \cite{helicopter_poinsot,multi-injector}. Similarly, a rocket combustor will be made of hundreds, if not thousands, of individual injector elements \cite{yangoefelin,rocket_multi}. In other applications, such as furnaces, the size of the domain is nominally very large compared to the resolution needed to ensure accuracy of sub-filter models, although some LES calculations have been performed \cite{kronenburg_furnace,kempf_furnace}. Hence, using computational tools for such complex systems requires additional modeling innovations.  

As data collection and storage become cost-effective, there is an opportunity to use data to obtain information or insight into the physical system. In many physics application domains, a suite of models that are derived from data have begun to emerge, leading to multiple approaches for introducing such data into predictive tools. In data-driven modeling, existing data from canonical or full-scale combustion systems are used to devise reduced descriptions. In the context of automotive applications, the notion of continuous set of models has been introduced \cite{albrecht2007towards}. In some sense, this is similar to the UQ procedure described above, but the use of data can extend beyond calibrating models for physical sub-processes. For some applications, models can be directly obtained from the data. Below, their relevance to combustion applications is discussed.

\subsection{Black-box and grey-box models}

Apart from the use of data for fault tolerance \cite{zedda_aiaa,ogaji_riti_ieee}, one of the most common combustion applications is the direct use of data to develop input-output maps. For instance, by varying combustor operating conditions and obtaining sensor outputs, it is possible to construct an empirical relation between input parameters and the quantities of interest. Here, two different approaches can be followed. In the first approach, the combustion system is treated as a black-box with a set of actuators and output sensors. Then, different techniques can be used to develop either static input/output maps \cite{siemens_ann,NN_diagnosis} or system-identification based dynamical systems (for instance, \cite{ghoniem,AI_combustion}). The former set of tools is used for predicting macroscopic quantities such as emissions or exit temperature, while the latter set is useful for detecting oscillatory patterns such as thermoacoustic instabilities. 

\begin{figure}
\center
\includegraphics[width=0.35\textwidth,trim={0cm 0cm 0cm 0cm},clip]{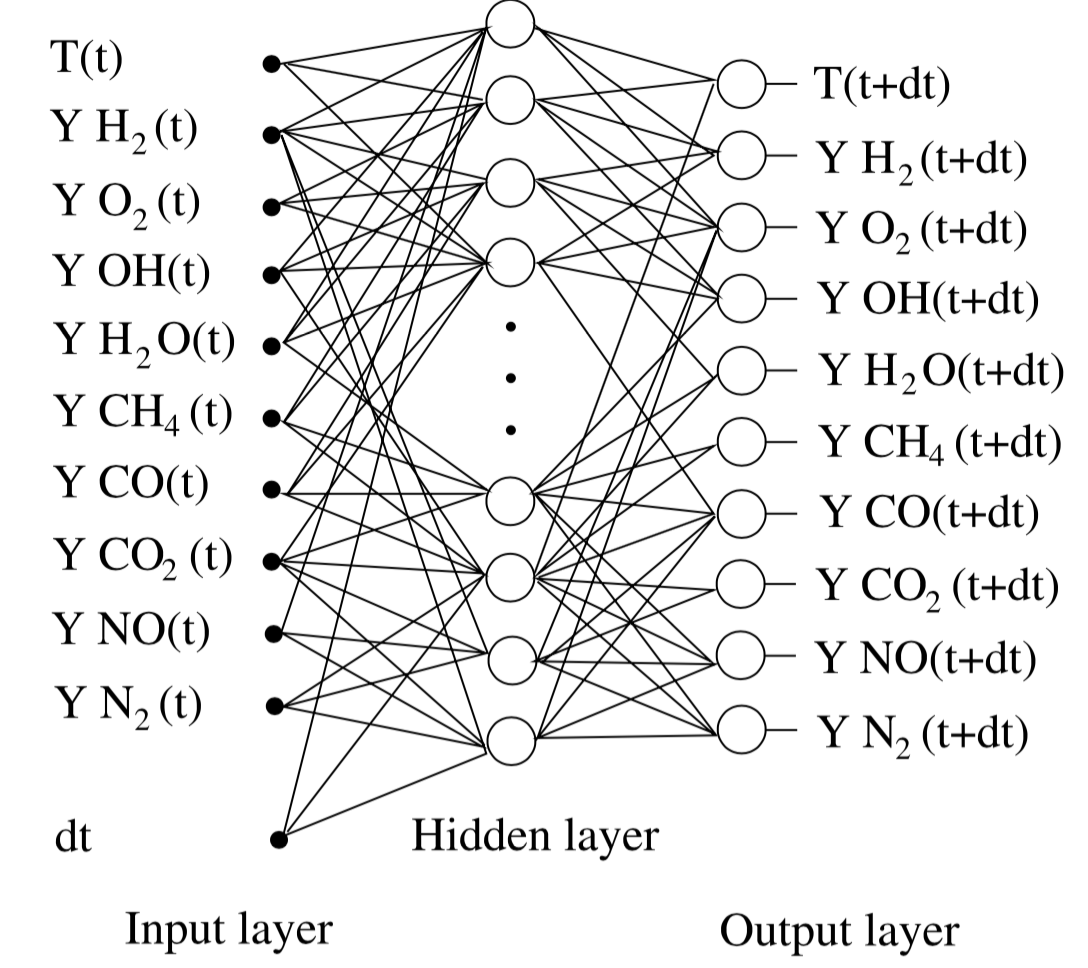}
\caption{Application of ANN to estimating chemical reaction rates. Figure shows structure of a multilayer perceptron employed to tabulate stiff reaction rates. Reproduced from Ref.~\cite{blascoSOM}.}
\label{fig:ann}
\end{figure}

In reacting CFD simulations, the static approach has been introduced with neural network models, primarily for modeling reaction rates. The advantages of the artificial neural network (ANN) are their ability to tabulate non-linear functions using minimal memory storage and perform smooth interpolation between the training points.  For instance, ANNs have been used to tabulate and evaluate reaction rates \cite{NN_chemistry_masri1995,blascoSOM,chemistryTabulation} and other model terms \cite{LEM_ANN_menon,NN_kempf,optimalANN_ihme}. Figure~\ref{fig:ann} shows a typical example of the information tabulated by these ANNs. In the second approach, called the grey-box modeling approach, simplified representations of the physical system are used to accelerate the computations. For instance, flame-transfer functions \cite{ftf_review} can be used to introduce the notion of flame response into the black-box models. {Note that applications of input-output map are not limited to predicting physical quantities, but can also inform parameters for control algorithms. Such iterative learning techniques have been used in the context of automotive applications~\cite{malikopoulos2007learning,zweigel2015iterative}.}

\begin{figure}[h]
\center
%\includegraphics[width=0.49\textwidth,trim={0cm 1cm 0cm 0cm},clip]{./Figures/ERN2.png}
%\caption{Example of a combustor divided into a network of several reactors. The algorithm that divides the combustor uses the flow configuration. Reproduced From Ref.~\cite{ern2}.}
\includegraphics[width=0.49\textwidth,trim={0cm 1.1cm 0cm 0cm},clip]{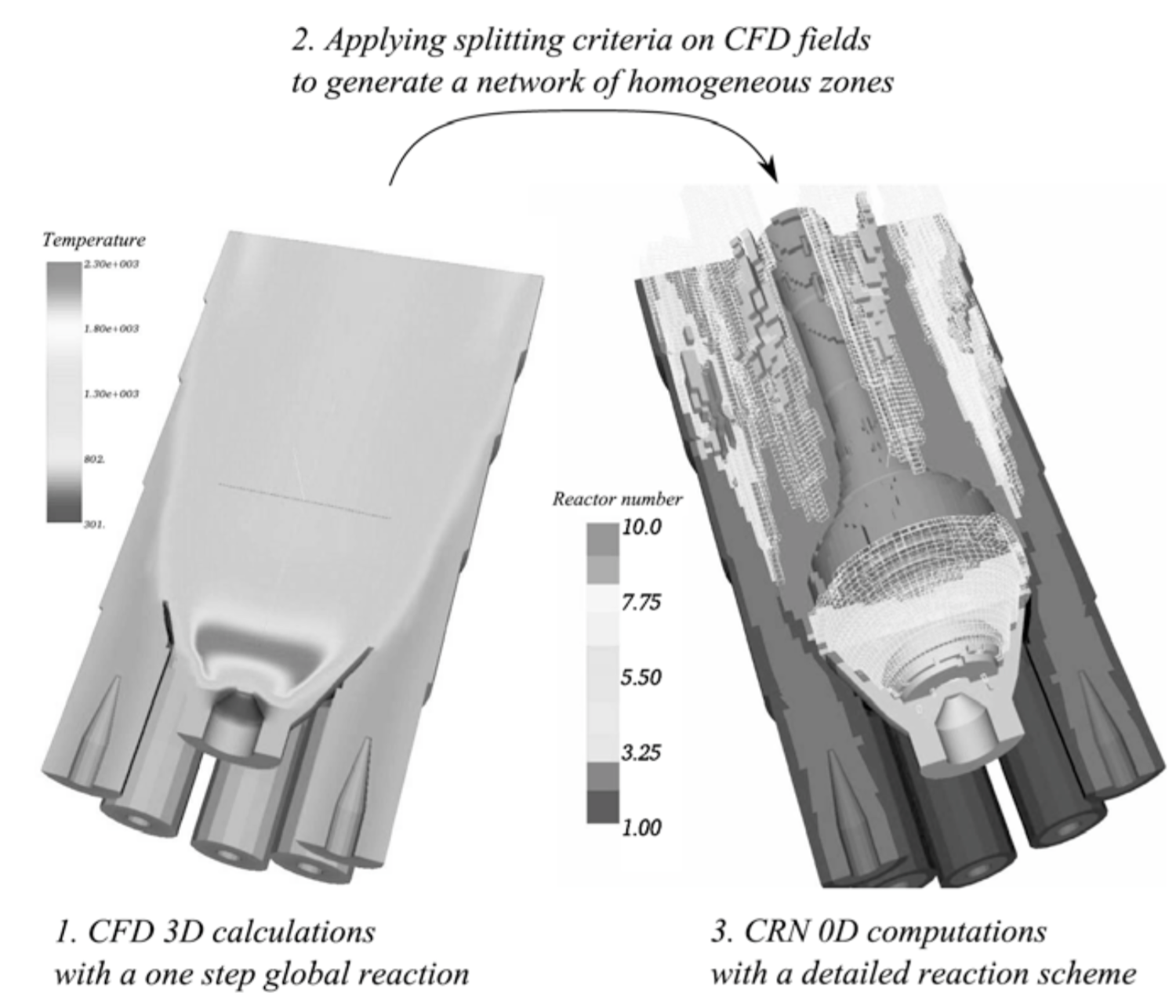}
\caption{Illustration of the process of building reactor network from a CFD result. Reproduced from Ref.~\cite{ecn-nox3}.}
\label{fig:ecn}
\end{figure}

Another interesting approach is the use of network models (for instance, \cite{ecn_algorithm,nox-ec1,nox-ecn2}). Here, the flow domain is decomposed into a set of well-stirred or other simple reactors that are connected. An application of this approach is to simulate only the flow field using high-fidelity tools to construct the reactor network. Then, detailed chemical kinetics can be used to simulate other quantities such as NOx \cite{nox-ec1,nox-ecn2,nox-ecn3} using this network model, which is nominally less expensive. Figure~\ref{fig:ecn} shows an example of the reactor network model constructed by a baseline CFD simulation of the flow field. These techniques are particularly useful for large and complex systems, such as furnaces or multi-injector gas turbines. In other words, partial information obtained from simulations is augmented by the network approach. Such models can be considered as the first step towards the set of data-driven reduced order models to be discussed below. 

\subsection{Reduced-order models}

Reduced-order models (ROMs) encompass a spectrum of methods and tools to simulate complex systems. As the name implies, the goal of ROMs is to reproduce specific characteristics of the combustion system at a highly-reduced computational cost. Since the development is based on the available data, ROMs are highly system-dependent while physics-based models attempt to describe a broad range of applications and conditions. The network model described above can be seen as physics-based ROM as well. Although different ROM approaches can be defined \cite{willcox_survey}, this section focuses on projection-based methods, primarily due to prior use in combustion as well as their utility in incorporating high-fidelity data.

\subsubsection{Projection-based ROMs}

A number of different methods fall under this category, but proper orthogonal decomposition (POD) \cite{lumleyholmes} has been commonly used in combustion applications (for instance, \cite{pod_comb2,pod_comb1}). Here, the variables of interest are expressed in terms of a set of basis functions with time-varying coefficients,
\begin{equation}
    u({\bf x},t) = \sum_{N_b} a_i(t)\Phi_i({\bf x}),
\end{equation}
where $N_b$ is the number of basis functions and $a_i$ is the modal coefficient of the $i$-th basis function $\Phi_i$. In many applications, the flow is decomposed into mean and fluctuating variables, and the expansion is applied only to the fluctuating component (for instance, \cite{steinberg,yang_pod}).{This type of expansion is similar to Fourier-series representation, with the difference that the basis function itself is computed from data}. Here, the data are snapshots of the flow field (at different time-steps) obtained from experiments or full-scale simulations, and basis functions are computed by formulating an eigenvalue problem \cite{sirovich1987turbulence,lumleyholmes,holmesbook}. Given $m$ snapshots of data at different times, an equal number of basis functions can be generated. Each basis function is also associated with an ``energy" based on a corresponding eigenvalue. In order to develop a reduced formulation, the basis functions with the highest $N_b$ energies are retained. This presupposes that for the ROM to be cost-efficient, the energy of the snapshots should be contained in a small number of modes. Alternatively, the choice of modes can be based on their impact on the system \cite{BalancedPOD,willcox_bpod} or other QoIs \cite{GoalROM}.

While the POD approach (and other such decomposition techniques) have been widely used to analyze data from LES and experimental sources, direct evolution of the modal coefficients to develop a predictive model has not been widely considered. To apply these expansions as a ROM, the projection operation (termed Galerkin approach) has to be applied to the governing equations, which will yield evolution equations for the modal coefficients. While such equations are easier to build for linear systems, highly non-linear systems encountered in combustion applications require special care \cite{huang_merkle,llnlreport}. The problem arises because the cost of Galerkin projection for non-linear equations can be as large as for the high-dimensional system. Further, similar to the RANS or LES approaches, the effect of discarded modes on the retained modes needs to be accurately modeled \cite{san2013proper,aubry,balajewicz2012,balajewicz2013}. New methodologies show that it is possible to approximate the non-linear part of the high-dimensional term using an optimal subspace \cite{DEIM,GNAT}.{Such approaches have been shown to be computationally efficient, for instance when approximating chemical source terms \cite{nguyen2014model}}. However, the development of ROMs for combustion applications is an open field with limited prior research.

\subsubsection{Non-intrusive ROMs}

A key challenge in the use of ROMs as discussed above is the development of the set of equations for the evolution of the modal coefficients based on the governing equations. Non-intrusive ROMs seek to circumvent this problem by directly inferring the evolution model for the dynamical system. For instance, in the POD-radial basis function (POD-RBF) approach \cite{podrbf}, the evolution of the POD coefficients are represented as a mapping function that advances the POD coefficients from one time-step to the next through an algebraic functional form. 

More recently, techniques such as dynamic mode decomposition (DMD) \cite{schmid2010jfm, crowley} and cluster-based reduced order modeling (CROM) \cite{crom_kaiser} have been developed. In the DMD approach, it is assumed that a linear mapping connects the flow field at time $t$ and the next time increment $t+\Delta t$ as:
\[
{\bm v}_{t+\Delta t} = A {\bm v}_{t},
\]
where the mapping $A$ is approximately the same over a sample time $t \in [0,T]$. As pointed out by Schmid \cite{schmid2010jfm}, this is equivalent to assuming a tangent linear model for the dynamics. The matrix $A$ is obtained using snapshots of experimental or numerical data obtained from the system of interest. Flows with oscillations around a fixed point are particularly amenable to this approach, provided the data set is sampled at frequencies that can capture these fluctuations \cite{schmid2010jfm}. The DMD-based analysis of flow fields has been used with many experimental \cite{dmd_swirl_markovich2014} and numerical data sets \cite{motheau2014mixed,dmd_swirl_2014carlsson,grenga2018dynamic}, but its utility as a ROM has been sparsely explored \cite{kutz_dmdrom}. 

\begin{figure}
\center
\includegraphics[width=0.49\textwidth,trim={0cm 0cm 0cm 0cm},clip]{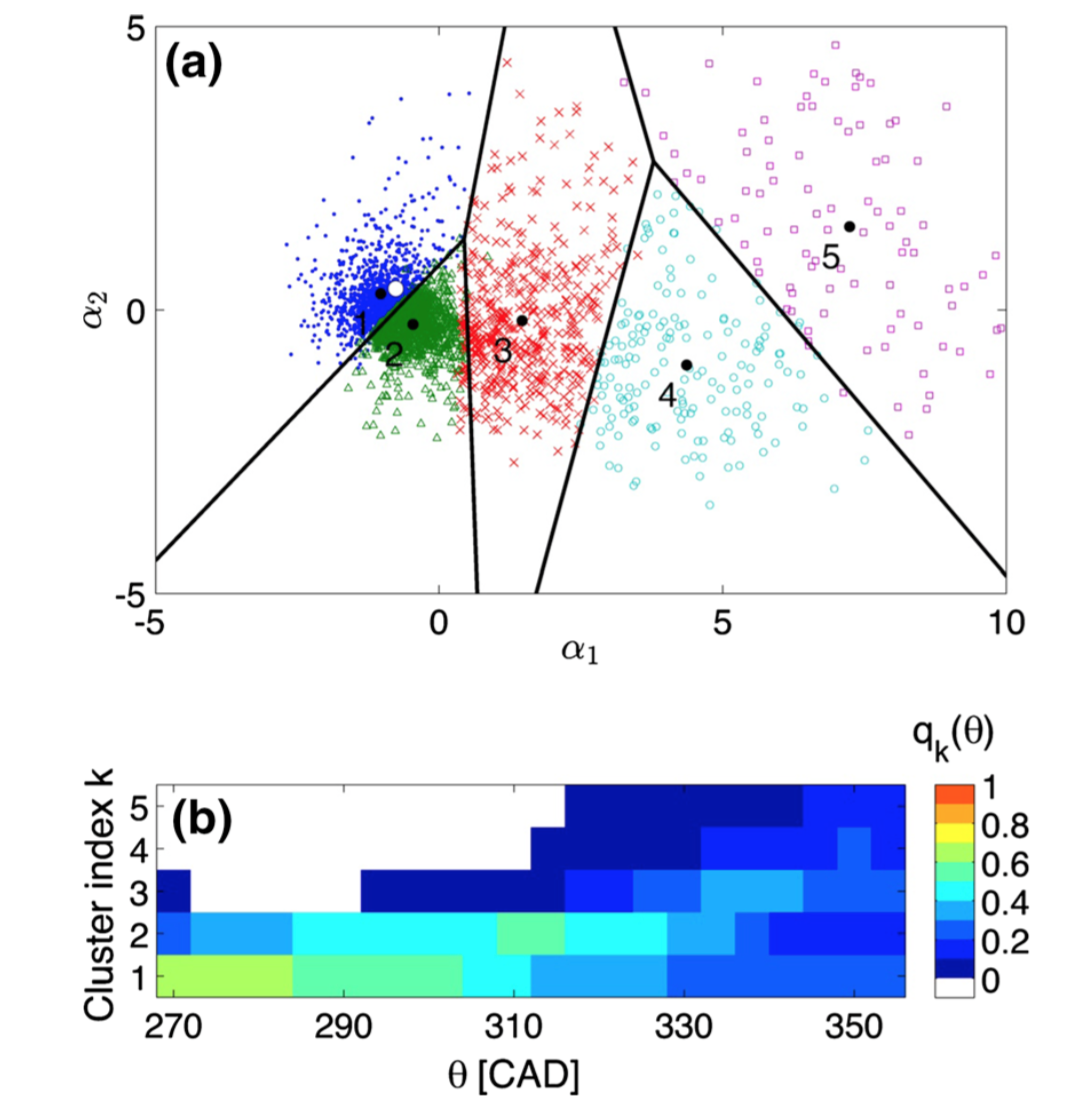}
\caption{Illustration of the CROM approach, which is based on clustering of data in phase-space, applied to data from internal combustion engine. (Top) Two dimensional Voronoi diagram showing the clustering process that groups experimental snapshots into ``nearby" regions in phase-space. (Bottom) Probability of encountering a particular cluster at different stages of the compression. Reproduced from Ref.~\cite{crom_ice}.}
\label{fig:crom}
\end{figure}

The alternative CROM technique takes a fundamentally different approach. All the ROM techniques discussed so far discount the statistical properties of the system. In the CROM approach, the phase-space of the system is discretized in terms of clusters, and the probability of transition between these clusters is estimated using the flow snapshots. Unlike the other ROM approaches, the CROM method seeks to approximate the change in phase-space density of the system \cite{crom_kaiser}. In this sense, it provides a statistical forecast as opposed to the solution to a deterministic ROM. Although this approach was developed only recently, application to predicting cycle-to-cycle variations in internal combustion engines has already been explored \cite{crom_ice}. Figure~\ref{fig:crom} shows how the flow field in a reciprocating engine can be classified among clusters. One feature that separates CROM from the POD-type techniques is its ability to approximate low-probability events in the system \cite{crom_kaiser}.  All the methods mentioned here are strongly dependent on the data that is being used. 

\section{Future paths for the use of computational models}
\label{sec:future}
The sections above showcase possible modeling trends. As discussed in Sec.~\ref{sec:intro}, there is a change in the application areas as well. In this section, the relevance of these tools to emerging applications is discussed. 

The discussions of the tool sets in the previous sections provide the simulation palette available to the end user. For each application, several of these tools will have to be used synchronously in order to achieve the end goal of optimal design or efficient operation. Here, existing and emerging paths are discussed.

\subsection{Design using multi-fidelity tools}

Design of complex systems cannot rely solely on high-fidelity tools, even if such tools are highly accurate. Design is often an iterative process which requires multiple model evaluations. Therefore, the need to speed up the iteration process (decrease the time to solution) may overwhelm the need for very high accuracy at some stage of the design. Multi-fidelity tools are an attempt to find a balance between both needs \cite{peherstorfer2016survey}. Here, models with different levels of fidelity (for instance, LES, RANS, and ROMs) will be used together to optimize a design for specific outcomes. For instance, high-fidelity models (e.g. LES) are used to calibrate lower-fidelity tools (RANS, ROMs).  There are several research challenges in such usage: 1) how to perform such calibration while minimizing the use of the high-fidelity solver, 2) how to decide on the fidelity of the model, 3) how to bridge information between the high and low fidelity solvers (see Ref.~\cite{peherstorfer2016survey} for an extensive review). In external aerodynamics, the use of multi-fidelity tools is already well-established \cite{robinson_multifid}, where detailed computations are supplemented by ROM, surrogate approaches, and lower-fidelity tools such as potential flow solvers \cite{xfoil}. Similar hierarchies have been formulated for reacting flows \cite{elderidgeaiaa,pod_moureau}. The underlying idea is that high-fidelity models are executed only for select designs, with lower-fidelity and ROMs used to interpolate between these design points. Within this framework, there is scope for including uncertainty quantification \cite{karniadakis} as well as data-based tools that incorporate experimental data.

\subsection{Digital twins and universes}

\begin{figure}
\center
\includegraphics[width=0.49\textwidth,trim={0cm 0cm 0cm 0cm},clip]{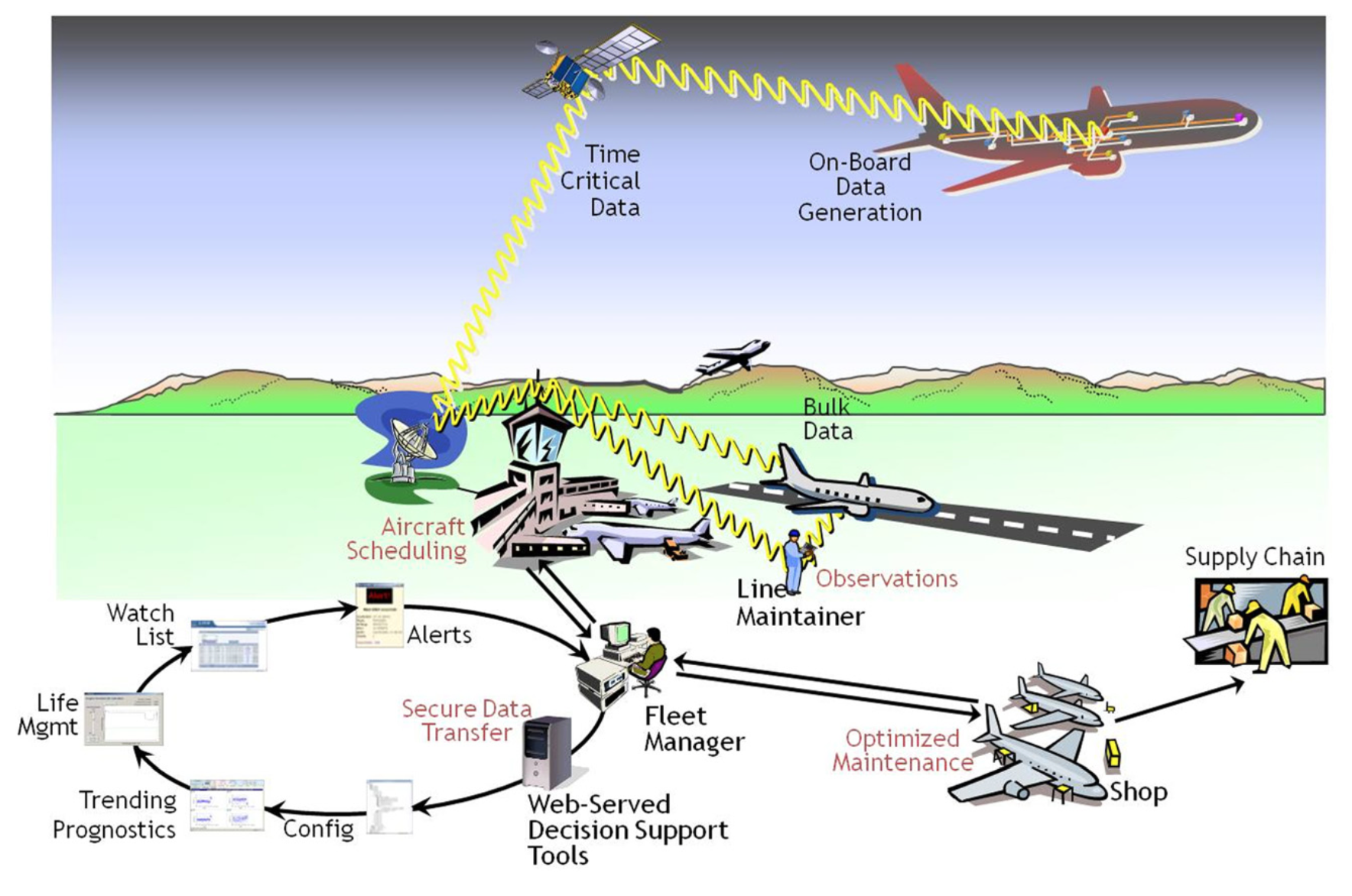}
\includegraphics[width=0.49\textwidth,trim={0cm 0cm 0cm 0cm},clip]{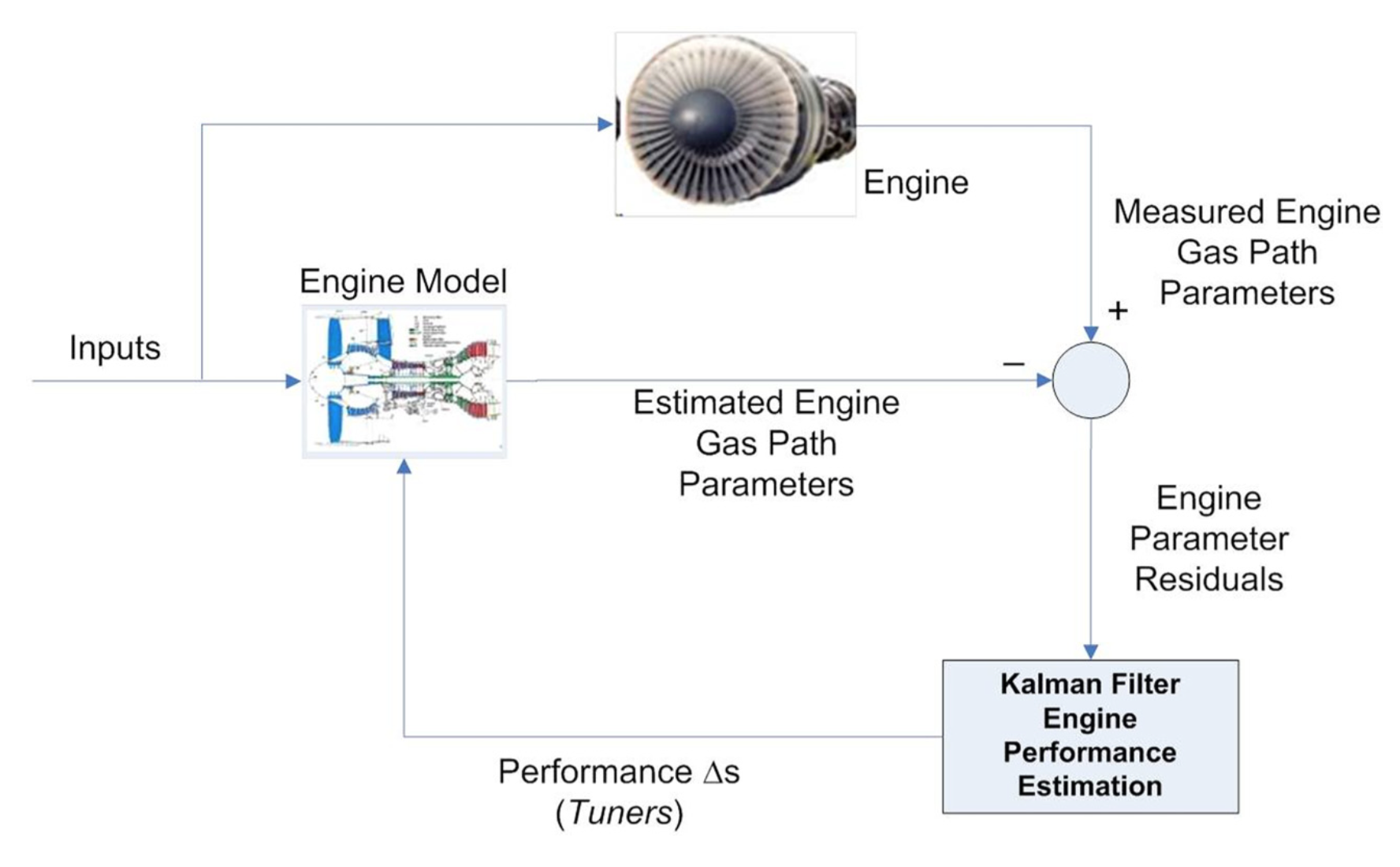}
\caption{Top: Illustration of a possible operational scheme for maintenance. Bottom: Example of engine management plan based on a model. Reproduced from Ref.~\cite{dtgt1}.}
\label{fig:dt}
\end{figure}

With the increase in importance of MRO, an emerging concept is the so-called digital twin \cite{dtwin1,dtwin2}. Modern manufacturing methods, especially for gas turbines, are highly digitized with an electronic inventory of even the smallest production detail \cite{dtwin2}. Further, the proliferation of cheaper and robust sensors combined with effective data collection ensures that the performance of physical assets can be tracked in real-time \cite{dtgt2}. An interesting use of this trove of information is the construction of a virtual copy of the physical system. This virtual asset can then be simulated using the suite of models and frameworks discussed in the previous sections. More importantly, the performance of the model and the physical system can be continuously checked, and when variations occur, a more in-depth analysis of the simulation results could be used to determine the cause of the deviation \cite{dtgt1}. Further, the digital twin can be subjected to different operational scenarios, and the future performance of the system assessed. Figure~\ref{fig:dt} illustrates how digital twins can be used to manage aircrafts. When used with a fleet of such twins \cite{dtgt2}, a digital universe can be developed. These forecasting capabilities are particularly useful for detecting performance drops, troubleshooting and predicting time to maintenance or down-time for stationary gas turbines. The development of such virtual assets is happening across multiple industries, and is becoming a dominant topic for combustion applications. As a recent example, the 2018 theme of the ASME IGTI meeting is ``Maintenance, Repair and Overhaul in the Light of Digitalization" \cite{asme_igti}. An important requirement for digital twins is the ability to forecast in near-real time. The simulation infrastructure should utilize multi-fidelity tools, but should also be able to assimilate sensor data efficiently.

\subsection{Model-based determination of failure events}

An extension of the above uses of computation is the ultimate goal of design: to minimize unforeseen failures. For many practical combustion systems, device failure in the form of loss of propulsion or structural damage due to excessive thermal loads can lead to catastrophic consequences. Hence, the design space is often restricted due to high margins required for safety. Unsurprisingly, gains in efficiency or emissions reductions are feasible by reducing these margins while ensuring safe and reliable operation. However, ensuring robustness of the system, especially for complex systems with non-linear behavior, is extremely difficult.

Traditionally, the issue of device failure has been studied as part of life-cycle analysis \cite{statfail}, where statistical tools are used to estimate time to failure. However, such approaches are inherently limiting since a) they require a failure event to be observed as part of the testing; b) such a failure is reproducible only using external stimuli so as to generate sufficient number of events for statistical measures; c) observed failures are the only type of failure events possible. It is this last aspect that leads to the issue of risk with complex systems. Since combustion systems are highly complex, failure initiates as a result of different mechanisms. Even full-scale testing may not be able reproduce such processes. For instance, ignitors for aircraft engines are designed to minimize time to successful flame stabilization. However, due to the chaotic flow environment, an upper bound on this time cannot be established purely by testing. 

In the recent past, there has been a growth in interest in the use of computational models to assess risk \cite{giardina2006direct,courantguy,tsapsis,malik_classification}. The goal here is to use detailed computational models with specifically designed methods to reliably construct extreme events. For turbulent combustion applications, there exists a number of philosophical and technical issues. From a simulation point of view, failure events are not all created equal and are caused by different sources. To address the source of risk, Hassanaly et al.~\cite{malik_classification} developed a classification of failure events of interest to combustion applications. They described three type of failures. The first type (type I) is a deterministic failure due to a change in macroscopic conditions, for example, the stall of an airfoil due a change in the angle of attack. The second failure type (type II) is due to uncertain initial conditions. This is important in the context of time-constrained system, such as ignition in a cold engine. The third type of failure is due to an uncertainty about the physical states that the device can exhibit. Here, failure can arise because of the slow dynamics of the system (type III-A), an external shock applied on the system (type III-B), or a modification of the system itself (type III-C). For such problems, critical states are obtained via a process of exploration rather than observation. Hence, efficient algorithms to explore high-dimensional phase-spaces are needed \cite{bouchet,malik_classification}, which in itself is a tremendous scientific challenge. Here again, combustion systems are amenable to simplifications due to the role of thermodynamic constraints, which could be used to limit the dimensions of the state-space. This approach is similar to that used for UQ (Sec.~\ref{sec:uq}), but requires specialized computing and modeling infrastructure. 

\begin{figure}
\center
\includegraphics[width=0.49\textwidth,trim={0cm 0cm 0cm 0cm},clip]{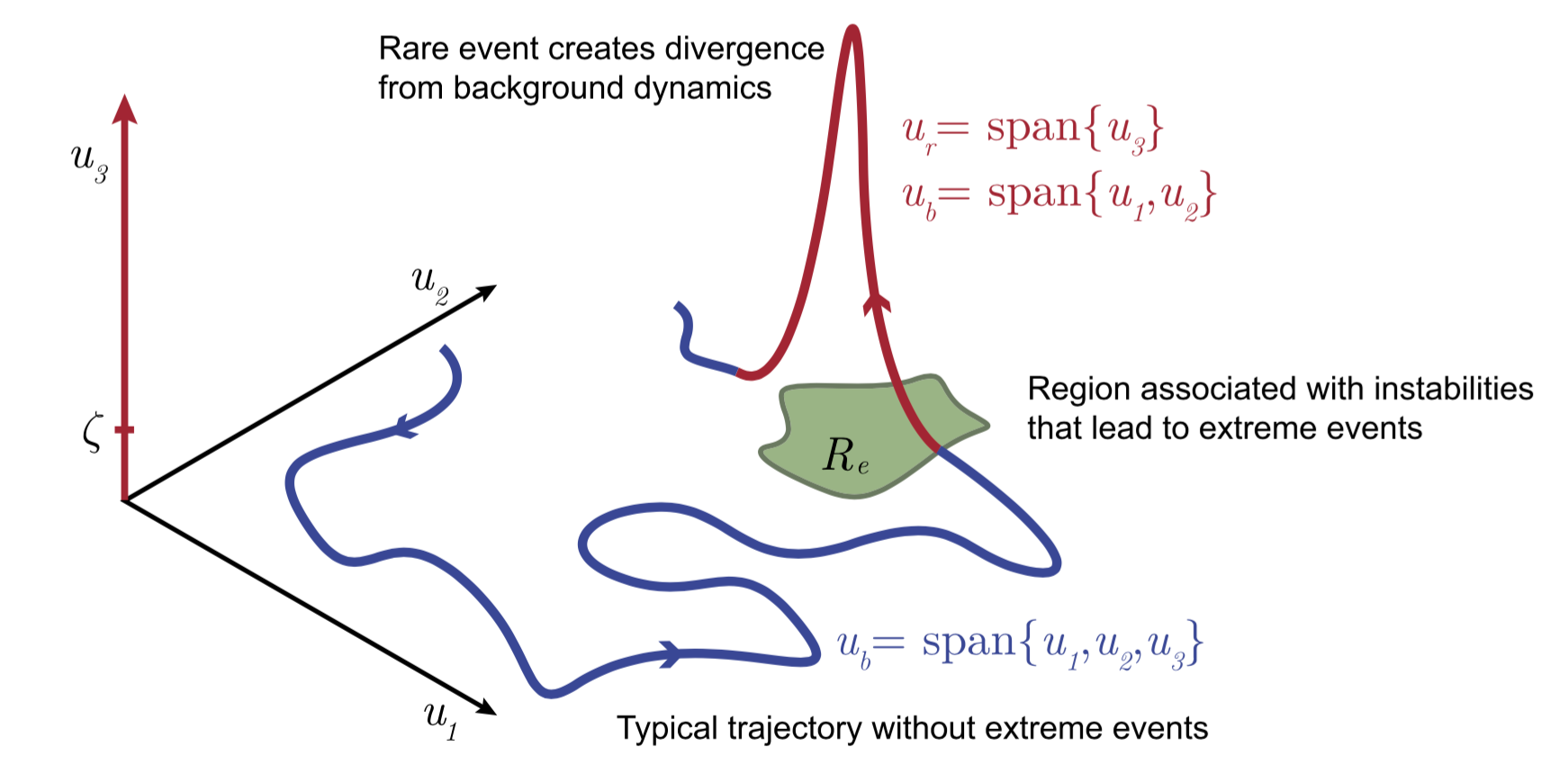}
\caption{Illustration of the path to instability. A potentially small region of phase-space (traversed by a small number of trajectories) can amplify instabilities and create a catastrophic event. Reproduced from Ref.~\cite{sapsis_instab}.}
\label{fig:risk}
\end{figure}

One critical limitation that is identified from \cite{malik_classification} is due to the statistical view of turbulence and combustion. As mentioned in Sec.~\ref{sec:current}, even an unsteady approach such as LES, when cast in a rigorous mathematical framework, yields only average paths in phase space \cite{langford_moser,pope_self}. Instead, extreme events are caused by rare trajectories in this phase-space which are associated with the tails of the probability density function of states observed for a particular system. The importance of extreme trajectories is illustrated on Fig.~\ref{fig:risk}. Consequently, sub-filter models and other aspects of the simulation that only seek such average trajectories \cite{langford_moser,optimal_moser, adrian} cannot capture these processes. To simulate such extreme paths, it is necessary to dispense with the statistical view and approach the problem using a dynamical systems (DS) point of view. During 1980s-1990s, the DS approach was widely studied for turbulence \cite{holmesbook}, predominantly as a method for model reduction. The same approach is remarkably suited for studying rare trajectories. Such advances are already used in short term weather forecasting \cite{kalnay,europeanprediction}. Recently, this DS perspective has been extended to turbulent combustion by Hassanaly and Raman \cite{malik-proci}. The ability to assess risk will open the design space to novel propulsion concepts, and help increase operational flexibility.

\section{Conclusions and outlook}
\label{sec:outlook}

The field of combustion modeling has evolved substantially since the start of this century, with LES-based modeling providing giant leaps in predictive accuracy. {However, even as changes are being made to combustor designs and their operational regimes, there is a saturation in the nature of LES models}. At the same time, the domain of applicability of computational tools is expanding with the rapid digitization of combustion systems. In this new era, data is ubiquitous, and its collection and storage are the intense focus of a number of research fields. Computing is not solely driven by high-performance supercomputers. Since both data collection and low-power computing hardware have become prevalent, there exist new opportunities for combustion modeling and numerical simulations that were not available even a decade ago. This provides a future outlook for combustion science that extends beyond modeling of turbulence-chemistry interactions:

\begin{itemize}
\item \textbf{Use of computing power:} So far, higher computing power has been nominally used to extend the details of the model in order to increase predictive accuracy. While this aspect will remain important, other uses of computing power should be considered. For instance, ensemble calculations for UQ or for predicting rare events. At the same time, low-power computing that is available on board combustion devices can be used to perform preliminary evaluation of system performance. In this sense, continually updated but low-fidelity models or on-board learning algorithms that adapt to the specific device could be formulated.
\item \textbf{Use of data:} The dominant force in combustion research is likely to be the availability of data, not just from experiments and DNS, but from fleets of operational devices. The ability to analyze such large volumes of data, either to improve performance or forecast device problems, will add tremendous value. Here, there is a danger that the use of data remains in the black-box mode, where the physics of combustors are neglected in favor of a purely data-driven approach. The ability to incorporate physics into data-based modeling is a necessary research direction. In this regard, the use of ROMs provides an opportunity for physics-driven modeling.
\item \textbf{Leveraging digitalization:} This aspect combines the above two topics. The availability of sensor data and the ability to perform small-scale on-board computations opens new paths for combustion modeling. In particular, processed data could be used to continually update models similar to software updates. In this case, both high-fidelity computations (LES, DNS) and experimental sciences can be used more creatively, to provide in-depth analysis of emerging operational scenarios. 
\end{itemize}

In a broad sense, future modeling can be categorized into three classes based on the availability of data: a) data-sufficient problems, where sufficient experimental data and/or high-fidelity computational tools exist such that most design variations and associated quantities of interest can be evaluated at reasonable cost. In particular, the experimental capabilities are well developed such that accurate measurements of these quantities can be made. Such data-sufficient problems are particularly useful for incremental design changes; b) data-rich problems, where digitization has led to continuous data streams that provide a wealth of sensor data from functioning devices, but only provide a partial view of the state of the device. The modeling challenges in this context are vastly different from the data-sufficient environment, and require fast-execution models that leverage the availability of data;c) data-poor problems, where neither experiments nor digitization can provide the data needed to extract useful insights or models. This includes the occurrence of low probability events or exploration of extreme operational environments with limited knowledge of the associated physics.

This discussion, hopefully, demonstrates that turbulent combustion and numerical modeling are at a crossroads. The continued success of this scientific community relies on our ability to adapt to the fast-paced changes sweeping the industrial world, and more broadly, society itself. The monolithic focus on LES or high-fidelity computational tools is neither sustainable nor fully relevant in the future. There is a clear need for introspection into current practices and objectives as well as a need to reach out to other research disciplines that have made considerable advances in related topics.

\section*{Acknowledgements}

The authors would like to acknowledge financial support from AFOSR through grant FA9550-15-1-0378 with Dr.\,Chiping Li as program manager, and grant FA9550-16-1-0309 with Dr.\,Jean-Luc Cambier as program manager.

%{\color{red} Malik - your section}

\section*{References}

\bibliographystyle{science.bst}

\end{document}